\documentclass[aps,prd,showpacs,showkeys,amsmath,amssymb]{revtex4}
\usepackage[english]{babel}
\usepackage{dcolumn}
\usepackage{epsfig,float,subfigure,color}
\makeatletter

\newcommand{\beq}{\begin{equation}}
\newcommand{\eeq}{\end{equation}}
\newcommand{\bea}{\begin{eqnarray}}
\newcommand{\eea}{\end{eqnarray}}

\usepackage{amssymb,amsmath}

\begin{document}
\title {Dependence of the black-body force on spacetime geometry and topology}
\author{C. R. MUNIZ{${}^{a}$\footnote{E-mail:celiomuniz@yahoo.com}}, G. ALENCAR{${}^{b}$\footnote{E-email: geovamaciel@gmail.com}}, M. S. CUNHA{${}^{c}$\footnote{E-email: marcony.cunha@uece.br}}, R. R. LANDIM{${}^{b}$\footnote{E-email: renan@fisica.ufc.br}}, R. N. COSTA FILHO{${}^{b}$}\footnote{E-mail: rai@fisica.ufc.br }}
\affiliation{
${}^{a}$\!Universidade Estadual do Cear\'a, Faculdade de Educa\c c\~ao, Ci\^encias e Letras de Iguatu, Rua Deocleciano Lima Verde, s/n, Iguatu-CE, Brazil. \\  ${}^{\!b}$\!\!\! Departamento de F\'{\i}sica, Universidade Federal do Cear\'{a}, Caixa Postal 6030, Campus do Pici, 60455-760, Fortaleza-CE, Brazil.\\ ${}^{c}$Grupo de F\'isica Te\'orica (GFT), Universidade Estadual do Cear\'a, 60714-903, Fortaleza-CE, Brazil.
}
\begin{abstract}
\noindent In this paper we compute the corrections to the black-body force (BBF) potential due to spacetime geometry and topology. This recently discovered attractive force on neutral atoms is caused by the thermal radiation emitted from black bodies and here we investigate it in relativistic gravitational systems with spherical and cylindrical symmetries. For some astrophysical objects we find that the corrected black-body potential is greater than the flat case, showing that this kind of correction can be quite relevant when curved spaces are considered. Then we consider four cases: The Schwarzschild spacetime, the global monopole, the non-relativistic infinity cylinder and the static cosmic string. For the spherically symmetric case of a massive body, we find that two corrections appear: One due to the gravitational modification of the temperature and the other due to the modification of the solid angle subtended by the atom. We apply the found results to a typical neutron star and to the Sun. For the global monopole, the modification in the black-body potential is of topological nature and it is due to the central solid angle deficit that occurs in the spacetime generated by that object. In the cylindrical case, which is locally flat, no gravitational correction to the temperature exists, as in the global monopole case. However, we find the curious fact that the BBF depends on the topology of the spacetime through the modification of the azimuthal angle and therefore of the solid angle. For the static cosmic string we find that the force is null for the zero thickness case.
\end{abstract}
\maketitle
\section{Introduction}

In a recent letter, M. Sonnleitner {\it et al.} \cite{Marte} studied counterintuitive forces which arise between a black-body and neutral atoms or molecules - named black-body force (BBF)  - due to the energy associated with shifts in the absorption lines of their spectra. These shifts are induced by the so-called dynamical Stark effect, in which the incoherent electromagnetic waves generated from the heated black-body surface disturbs the electronic transitions in the atomic spectrum.

In the above mentioned article, the authors have found that attractive forces undoubtedly arise from shifts in the electronic transitions involving the $1s$ level of the hydrogen atom, and such forces would eventually be stronger than the repulsive force caused by the radiation pressure and the gravitational force due to the black-body mass. However, the energy associated to the spectral shift induced in other atom species could be positive, leading to repulsive forces. In the microscopic level, this effect can arise from other incoherent electromagnetic emissions, strongly dependent on the shape of the emitting surface. It is also noticeable in \cite{Marte} the emphasis put on the connection between the geometry of the bodies and the BBF, but the authors did not investigate the possible influences of both the topology and spacetime geometry on that force. In other words, those authors limited themselves to flat geometry with trivial topology.

Astrophysical and cosmological scenarios as well as some models in condensed matter physics offer possible investigation lines for the BBF. However, corrections to that force are necessary because even in a weak-field regime, in standard or alternative theories of gravitation, there are non-negligible modifications in the black-body radiation law caused by the gravity \cite{muniz, muniz2}, which must be taken into account for a complete understanding of the phenomena. This is another sense in which geometry is fundamental in understanding of the BBF.

It is worth noticing that the authors of Ref.\cite{Marte} show that the BBF depends on both the temperature and the solid angle, which are modified by the spacetime geometry. Therefore this work focus on the effects due to both the spacetime geometry and topology on the BBF. This is considered for static and spherical gravitational sources as well as for infinitely long cylinders, including those ones like a cosmic string. We compute these corrections for the black-body potential and study their behavior comparing them with the flat case, obtained in \cite{Marte}. In the case of the spherical symmetry we initially will particularize for the Schwarzschild solution, making equally manifest the role of the local spacetime geometry in the black-body potential. We will numerically calculate this corrected potential regarding typical neutron stars and the Sun, comparing them with the situation in the flat spacetime. We will also calculate the black-body potential due to the global monopole, whose major feature is the deficit in the central solid angle of the metric generated by that object, modifying its topology \cite{Eugenio,Shi}. We will show that the difference from the flat spacetime black-body potential, different from the previous situation, grows with the distance to the source center.

Regarding the cylindrical geometry, whose BBF is initially calculated for the infinitely long Newtonian cylinder, an angular (deficit) parameter $\nu$ is appropriately introduced in order to investigate effects of the spacetime topology on that force. Such a parameter multiplies the azimuthal angle entering in the calculation of the solid angle subtended by the atom, and it characterizes the global properties of the cosmic string spacetime, which presents a locally flat geometry, but with global curvature. This feature generates interesting effects, as self-interaction on particles and dipoles \cite{anderson,celio3} and gravitational lens \cite{vilenkin}.

The cosmic string spacetime has a non-trivial topology which is encoded in the aforementioned angular parameter given by $\nu=(1-4G\mu)^{-1}$ \cite{vilenkin}, where $G$ is the Newtonian constant of gravity and $\mu$ is the linear mass density of the string. The more recent observations of the cosmic microwave background (CMB) provide an upper bound on dimensionless parameter $G\mu$ of the order of $10^{-7}$ \cite{Arde}. It is worth emphasize that cosmic strings can have played an important role in the primitive Universe, and it is known that in its very early ages the temperature was extremely hot \cite{peebles}; thus it is natural thinking that those objects were in constant interaction with the thermal radiation that bathed the Universe. Indeed, the effects of the thermal equilibrium of cosmic strings with radiation were analyzed in \cite{davies}. Although cosmic strings are still subjects of theoretical speculation, a concrete realization of them occurs in some condensed matter systems. The static cosmic strings are emulated as topological defects termed disclinations, which lie within crystalline structures, including liquid crystals in the nematic phase \cite{Lozar,Satiro}.

The paper is organized as follows. In section 2, we analyze the black-body potential on a neutral atom due to the radiation emitted by spherical sources, studying changes in this potential due to the gravitational field and the topology. In section 3, we consider cylindrical sources in the calculation of BBF, revealing the role of the topology in the production of it and, finally, in section 4 we discuss the results and close the paper.

\section{Correction to the black-body potential due to the spherical sources}

In this section we examine the influence of both spacetime local geometry and topology of spherical sources on the black-body potential, by analyzing the cases of a massive body and the global monopole.

\subsection{The static massive body}

As shown in Ref. \cite{Marte}, for the non-relativistic case the attractive potential associated to this force exerted on the neutral hydrogen atom in its fundamental state, placed at a distance $r$ from the center of a spherical black-body of radius $R$ at temperature $T$ is
\begin{equation}\label{potential}
V_{bb}(r)=-\frac{3\pi^3 (\kappa T)^4}{5 (\alpha_f m_e)^3}\frac{\Omega_a}{4\pi},
\end{equation}
where $\alpha_f$ is the fine structure constant, $m_e$ is the mass of the electron, and $\Omega_a$ is the solid angle subtended by the atom. About the above expression some points are worth comment.  As pointed in Ref. \cite{Marte}, for the hydrogen atom in an isotropic thermal bath, the dynamical Stark effect will
mostly cause a negative energy shift for the ground state. If the temperature of the thermal bath is $T$, this shift is given by
\begin{equation}\label{shift}
\Delta E_{1s}\approx -\frac{3\pi^3 (\kappa T)^4}{5 (\alpha_f m_e)^3}.
\end{equation}

Since the atom is in an isotropic thermal bath, the black-body field cannot have any directional effect on the
atom motion, inducing only friction and diffusion. Those authors then argue that if a finite source is considered, this produces a net force.
The point is that in this case, the atom does not see radiation coming from all directions, but only from a solid angle determined by the source.
Therefore, only a fraction of the energy shift in the isotropic bath would cause the BBF.  This fraction is given by the solid angle compared to $4\pi$, giving us  the final potential energy of Eq. (\ref{potential}).
 For the non-relativistic spherically symmetric case this solid angle is given by
\begin{equation}\label{solido}
\Omega_a=2\pi\left(1-\cos\theta\right)=2\pi\left(1-\frac{\sqrt{r^2-R^2}}{r}\right).
\end{equation}

From the above expression we can see that the radial dependence of the potential comes only from the solid angle. However, when we consider curved static spacetimes, the effective temperature will also depend on the coordinates via Tolman temperature \cite{Tolman:1930zza,Tolman:1930ona}. Therefore, in order to find corrections to the BBF two aspects must be considered: the first is the curvature correction to the temperature and the second is the modification in the solid angle determined by the spacetime curvature. This is the general procedure that we will follow throughout this manuscript.

The static and spherically symmetric exterior solution of Einstein's equation of General Relativity is initially considered, which is given by the metric
\begin{equation}\label{metric}
ds^2=\left(1-\frac{r_s}{r}\right)c^2dt^2-\left(1-\frac{r_s}{r}\right)^{-1}dr^2-r^2d\Omega^2,
\end{equation}
where $r_s=2GM/c^2$. As said above, the spacetime modification to the temperature is given by the Tolman temperature. At the position of the atom this is given by
\begin{equation}\label{Teff}
T_{eff}=\frac{T}{\sqrt{g_{tt}}}=\frac{T}{\sqrt{ 1-\frac{r_s}{r}}} ,
\end{equation}
where $T$ is the temperature of the isotropic thermal bath, as measured by an observer at the infinity. The fact that the temperature depends on the radial coordinate will contribute to modify the BBF.

Now we must consider the modification in the solid angle due to the spacetime curvature which occurs in the system investigated here. This is done by considering the change $\cos\theta\to\cos\tilde{\theta}$ in Eq. (\ref{solido}). According to \cite{Rindler:2006km}, and using the fact that in the flat limit
$$r^2\frac{d\varphi^2}{dr^2}=\tan^2\theta=\frac{R^2}{r^2-R^2},$$ the correct expression in Schwarzschild spacetime is given by
\begin{equation}\label{cos2}
\cos^{2}\!\tilde{\theta}=\frac{r^{2}-\frac{A(r)}{A(R)}R^{2}}{r^{2}}
\end{equation}
where $A(r)=1-r_s/{r}$. Thus we obtain the modification in the solid angle subtended by the atom, which is given by
\begin{equation}\label{solidomodificado}
\Omega_a=2\pi\left(1-\cos\tilde{\theta}\right)=2\pi\left(1-\frac{\sqrt{r^2-\frac{A(r)}{A(R)} R^2}}{r}\right).
\end{equation}
Notice that the above expression reproduces the Euclidean solid angle for $r_s\rightarrow 0$.
Plugging Eq. (\ref{Teff}) and Eq. (\ref{solidomodificado}) into Eq. (\ref{potential}) we get the final corrected expression for the black-body potential
\begin{eqnarray}\label{19}
V_{bb}(r)=-\frac{3\pi^3 (\kappa T)^4}{10 (\alpha_f m_e)^3}\frac{1}{A^2(r)}\left(1-\frac{\sqrt{r^2-\frac{A(r)}{A(R)} R^2}}{r}\right).
\end{eqnarray}
This potential corresponds to a force that also drops with the cube-inverse at large distances and is infinity when $r=R$, as in the flat spacetime case.

\begin{figure}[h!]
\centering
\includegraphics[width=8.9cm]{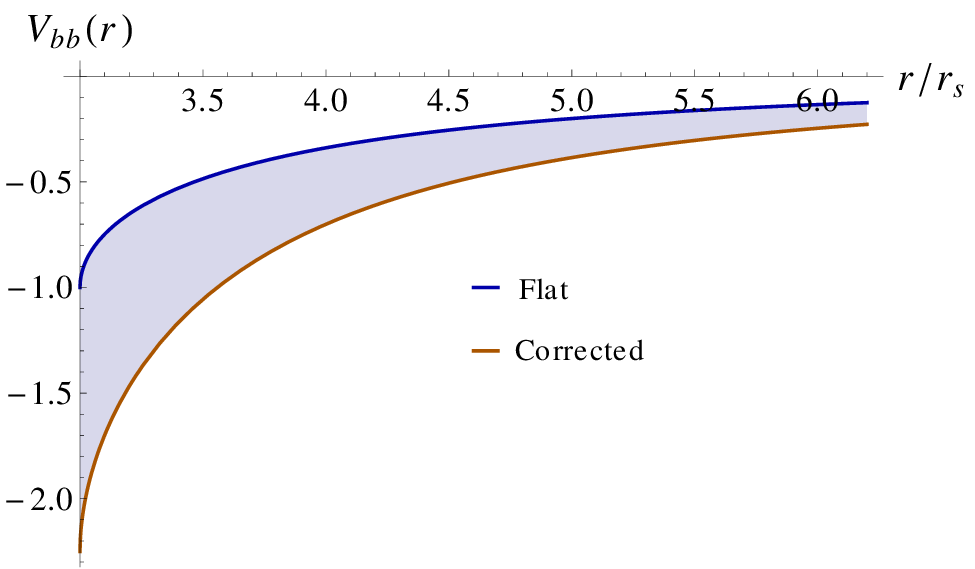}
\includegraphics[width=8.9cm]{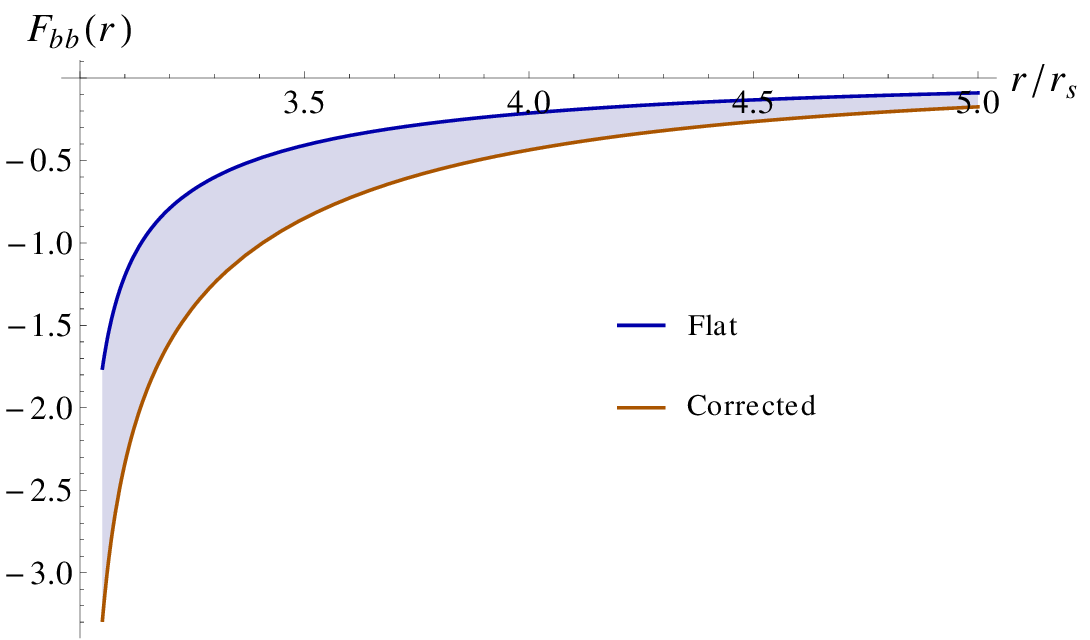}
\caption{The temperature-independent part of the black-body potential $V_{bb}$ and  force $F_{bb}$, gravitationally corrected and flat, both as functions of the ratio $r/r_s$, due to a massive spherical body with $R=3r_s$.}\label{1}
\end{figure}

The radial force derived from the black-body potential can be calculated via
\begin{equation} \label{radial force}
F_{bb}(r)=-\frac{dV_{bb}(r)}{d\ell}=-\frac{dV_{bb}(r)}{dr}\frac{dr}{d\ell},
\end{equation}
where $dr/d\ell=(1-r_S/r)^\frac{1}{2},$ according to the metric (\ref{metric}), with $\ell$ being the radial ruler distance \cite{Rindler:2006km}. The $V(r)$ asymptotic behaviour when $r$ becomes big is given by

In Fig. 1, we depict the graph of the relativistically corrected potential $V_{bb}$ over the temperature function $f(T)=\frac{3\pi^2 (\kappa T)^4}{10 (\alpha_f m_e)^3}$, against the same potential energy in the flat case, both depending on the ratio $r/r_s$, for a compact sphere with $R=3r_s$, which corresponds to the dimensions of a typical neutron star with about one solar mass. The difference between both the potentials is sensible, and the magnitude of the corrected one is considerably greater than the observed in the other case. At $r=4.5r_s$, the relative difference between the potentials is $(V_{bb}^{Corr.}-V_{bb}^{Flat})/V_{bb}^{Flat}\approx 100\%$, corresponding to a force approximately $108\%$ stronger. At shorter distances and for more compacts objects, the difference increases even more.

Notice that by considering the Sun, this reduction in the temperature-independent part of the black-body potential is {approximately 0.0007\% at $r=1.5R$}. For a quantum gravity correction to this force, see \cite{alencar}.


\subsection{The global monopole}

The case to be analyzed here is that of a global monopole, whose exterior metric is given by \cite{Eugenio,Shi}
\begin{equation}\label{globalmonopole}
ds^2=c^2dt^2-dr^2-\beta^2r^2d\Omega^2,
\end{equation}
 where $\beta^2 = 1 - 8 \pi G\,\eta_0^2 < 1$ is the parameter that gives the deficit in the central solid angle, modifying the spacetime topology. In order to calculate the BBF on the atom placed in this spacetime, we redefine the polar angle as $d\tilde{\Omega} = \beta d\Omega$, such that $\tilde{\Omega} = 4\pi \beta$ \footnote{Unfortunately, the corresponding Eqs. (\ref{polarangle})-(\ref{pot_inf}) are not correct in the EPL published paper and even in the EPL erratum. We apologize to readers for any inconvenience.}. The relation between the modified and non-modified angles at the source is defined as $2\pi\beta (1 - \cos\varphi) = 2\pi (1 - \cos\tilde{\varphi})$ or
 \begin{equation}
 \cos\tilde{\varphi} = 1 - \beta (1 - \cos\varphi) = 1 - \beta\left(1 - \frac{R}{r}\right). \label{polarangle}
 \end{equation}
 Since the relation $\cos\tilde{\theta} = \sin\tilde{\varphi}$ is valid, we can therefore write the correct expression for the solid angle subtended by the atom which is now
 \begin{equation} \label{EqOmega}
\tilde{\Omega}_a = 2\pi\left[1 -  \sqrt{\beta\left(1-\frac{R}{r}\right)\left(2 - \beta +\beta\,\frac{R}{r}\right)}\,\right].
 \end{equation}
If $\beta\rightarrow 1$ we obtain the BBF originally found in Ref. \cite{Marte}, since the redefinition of the polar angle makes the metric (\ref{globalmonopole}) locally flat in this limit. Expressing the new polar angle according to Eq. (\ref{polarangle}), we arrive at
 \begin{eqnarray}\label{Eq_potbb}
 V_{bb}(r)=-\frac{3\pi^{3}(\kappa T)^{4}}{10\,(\alpha_f m_{e})^{3}}\left[1-\sqrt{\beta\left(1-\frac{R}{r}\right)\left(2-\beta\left(1-\frac{R}{r}\right)\right)}\,\right]
\end{eqnarray}
\begin{figure}[hb!]
\centering
\includegraphics[width=8.9cm]{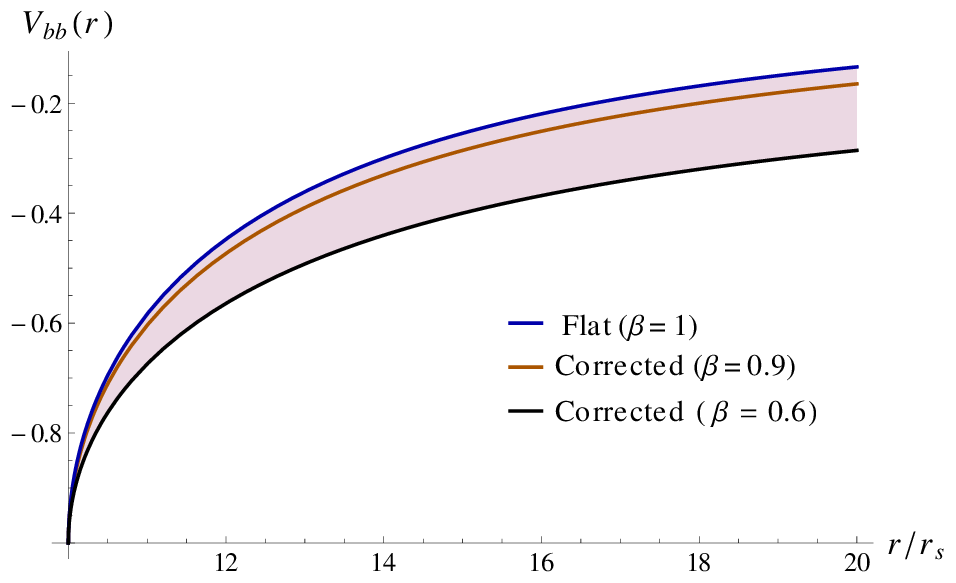}
\includegraphics[width=8.9cm]{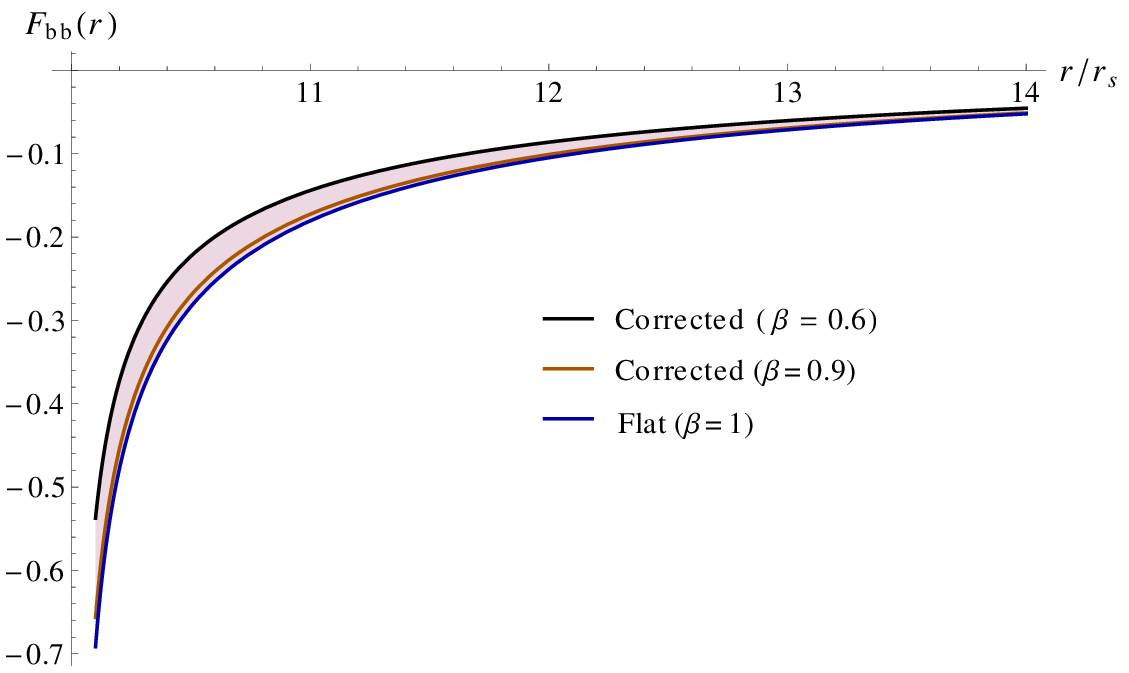}
\caption{The temperature-independent part of the black-body potential and force, $V_{bb}$ and $F_{bb}$, gravitationally corrected and flat, as a function of the radial coordinate, in kilometers, due to a global monopole. The $F_{bb}$ graphic is presented in a shorter region where it is more relevant.}\label{fig2}
\end{figure}
and therefore the force is written as
 \begin{eqnarray}\label{eq10}
F_{bb}(r)=-\frac{3\pi^{3}(\kappa T)^{4}}{10\,(\alpha\, m_{e})^{3}}~\frac{R\, \beta \left[1 - \beta \left(1 - \frac{R}{r}\right)\right]}{r^2\sqrt{\beta\left(1 - \frac{R}{r}\right) \left[2 -\beta \left(1 - \frac{R}{r}\right)\right]}}
\end{eqnarray}

In fig. \ref{fig2}, we plot the temperature-independent part of the potential and force for the flat and corrected cases. There we have used $\beta = 0.9$ and $\beta = 0.6$ cases in order to stress the difference between the flat ($\beta = 1$) and curved space. In all cases we take $R = 10\, km$. Additionally, notice that the corrected potential at the infinity tends to a non-null result, \textit{i.e.},
\begin{equation}
V_{bb}(\infty)=-\frac{3\pi^3 (\kappa T)^4}{10 (\alpha_f m_e)^3}\left(1-\sqrt{\beta(2-\beta)}\right).\label{pot_inf}
\end{equation}\\
\section{Correction to the blakc-body potential due to cylindrical sources}

In this section we will consider the BBF caused by systems with cylindrical symmetries. There are two cases to be considered here. One is the non-relativistic radiating infinity cylinder, which was not considered in the original Ref. \cite{Marte}. The other case considers the influence of the spacetime generated by a static cosmic string, with and without thickness.

\subsection{The non-massive infinity cylinder}

From the expression without gravity (\ref{potential}), the only change in the potential will be due to the solid angle. Supposing that the particle is at a distance $d$ of the rectangle center, thus the solid angle of the whole cylinder subtended by the atom is given by \cite{Khadjavi}
\begin{equation}\label{CylinderSolidAngle}
\Omega_a(d)=4\pi\arccos{\sqrt{\frac{1+\alpha^2+\beta^2}{(1+\alpha^2)(1+\beta^2)}}},
\end{equation}
where $\alpha=\textsl{a}/2d$, $\beta=\textsl{b}/2d$; $\textsl{a}$ and $\textsl{b}$ are the length and width of the rectangle, respectively. Taking an infinitely long cylinder, $\textsl{a}\rightarrow\infty$, and writing the width in terms of the cylinder radius, $R$, and of the distance of the particle from the cylinder symmetry axis, $r$, which coincides with the $z$ axis, we have
\begin{equation}\label{02}
\Omega_a(r)=4\pi \arcsin{\left(\frac{R}{r}\right)},
\end{equation}
with $r>R$. The $\arcsin(R/r)$ term is just the half-angle measured from the atom and which embraces the cylinder in the plane $z=0$. In the Fig. 3 we depict the neutral atom (A) close to the cylindrical black-body. According to it, the geometrical elements present in our analysis are: $\overline{AD}\equiv d$, $\overline{OA}\equiv r$, $\overline{BC}\equiv \textsl{b}$, $\overline{OB},\overline{OC}\equiv R$, and $\widehat{BOC}\equiv \varphi$ \text{(azimuthal or central angle)}.

\begin{figure}[h!]
\centering
\includegraphics[width=6cm]{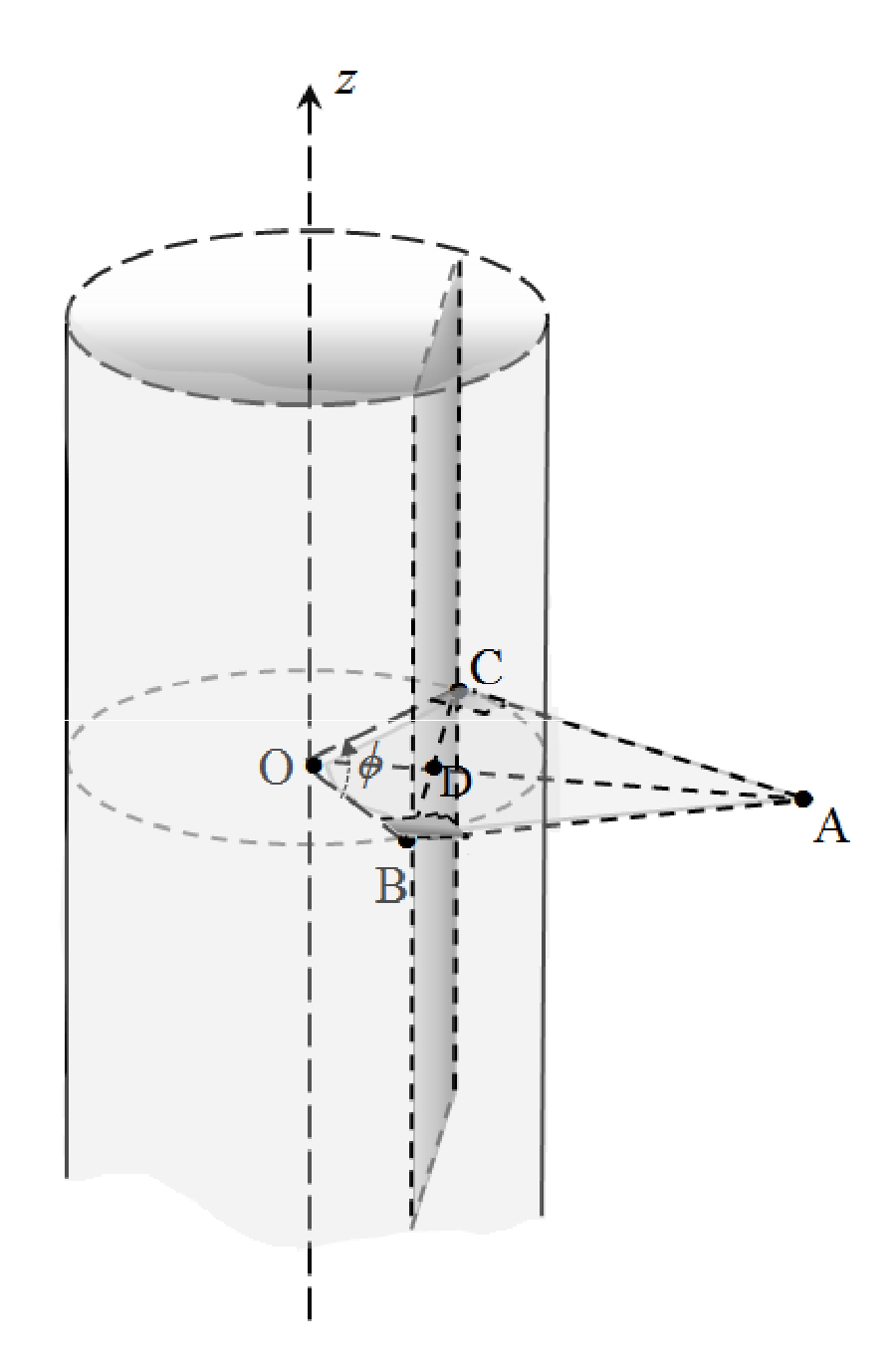}
\caption{Representation of the neutral atom (A) close to the cylindrical black-body.}
\end{figure}

We find that the attractive potential on the hydrogen atom in the ground state due to the cylindrical black-body at absolute temperature $T$ is given by \cite{Marte}
\begin{equation}\label{potentialNRC}
V_{bb}(r)=\Delta E_{1s}(T)\frac{\Omega_a(r)}{4\pi}=-\frac{3\pi^3(\kappa T)^4}{5\alpha_f^3m_e^3}\arcsin{\left(\frac{R}{r}\right)},
\end{equation}
where $\Delta E_{1s}(T)$ is the thermal shift in the absorption lines of the spectrum of the unexcited hydrogen atom. Thus the magnitude of the radial force exerted on this atom is given by
\begin{equation}\label{04}
F(r)=-\frac{3\pi^2(\kappa T)^4}{5\alpha_f^3m_e^3}\frac{R}{r^2\sqrt{1-R^2/r^2}},
\end{equation}
Notice that the BBF is attractive, falling with a square-inverse law at large distances (or $R\rightarrow0$). That is different from what happens in the spherical black-bodies where the force drops with the cubic-inverse law at those distances \cite{Marte}.

\subsection{The static cosmic string}
The fact that the cosmic string spacetime has a locally flat geometry but a non-trivial topology raises interesting points. First of all, due to the local flatness, the effective temperature is not changed and therefore the only possible correction comes from the solid angle like in the previous case. The interesting point is that the solid angle now depends on the deficit angular parameter $\nu$, incorporating then topological features in the BBF.

Let us model a static cosmic string, or its realization as a disclination in a liquid crystal, by the cylinder above considered. First, the conical geometry of the space around of the defect has to be incorporated. Such a geometry is described through the metric
\begin{equation}\label{05}
ds^2=-dt^2+dr^2+\frac{r^2}{\nu^2}d\varphi^2+dz^2.
\end{equation}
This expression describes the space around an ideal cosmic string of zero thickness, or the exterior region of that one endowed with an internal structure \cite{allen}. For this defect, $\nu>1$ (angular deficit parameter), and $0<\varphi\leq2\pi$. For a disclination, $\nu$ can also be between zero and one, meaning that we have an angular excess parameter. Then all we have to do is rewrite the half-angle from which the atom sees the cylinder, $\arcsin{(R/r)}$, in terms of the correspondent half central (azimuthal) angle, $\varphi/2=\arccos{(R/r)}$, such that
\begin{eqnarray}\label{06}
\arcsin{\left(\frac{R}{r}\right)}=\frac{\pi}{2}-\arccos{\left(\frac{R}{r}\right)},
\end{eqnarray}
and to insert the factor $\nu^{-1}$ which multiplies the azimuthal angles, according to the metric (\ref{05}). Thus the expression (\ref{potentialNRC}) becomes
\begin{equation}\label{Vnu}
V_{bb}(r)=-\frac{3\pi^3(\kappa T)^4}{5\alpha_f^3m_e^3}\left[\frac{\pi}{2}-\nu^{-1}\arccos{\left(\frac{R}{r}\right)}\right],
\end{equation}
and calculating the force via $F(r)=-dV_{bb}/dr$, we get
\begin{equation}\label{Fnu}
F(r)=\Delta E_{1s}(T)\,\frac{R}{r^2 \nu \sqrt{1-\frac{R^2}{r^2}}}.
\end{equation}
It is worth notice that, for the neutral hydrogen atom in the Minkowsky space ({\it i.e.}, when $\nu=1$) we have the expression (\ref{04}) again, and when one takes the limit $R\rightarrow 0$ valid for an ideal cosmic string, the BBF vanishes. Then, an important conclusion is that ideal cosmic strings do not exhibit the BBF behaviour.

For non-zero thickness and $R\ll r$, we can approximate Eq. (\ref{Fnu}) by
\begin{equation}\label{09}
F(r)\approx\Delta E_{1s}(T)\frac{R}{r^2 \nu},
\end{equation}
the force is attractive, due to the factor $\Delta E_{1s}$ to be negative for neutral hydrogen atoms, and it still falls off with the inverse square of the radial distance. Such a law is the same one that manages the self-force on neutral (or charged) massive particles situated in the spacetime of a Gott-Hiscock cosmic string, when the particle is far away from the string \cite{valdir}.

\section{Concluding Remarks}

We have studied the effects of the black-body radiation produced by spherical and cylindrical sources on a nearby neutral hydrogen atom in the ground state, which suffers, via dynamical Stark effect, a shift in its spectrum absorption lines producing a attractive force (BBF). This was made by analyzing geometrical and topological influences of their respective space-times on it.

Initially, we have calculated the effects due to a static and spherically symmetric gravitational background, starting by taking the Tolman temperature and substituting it in BBF general expression, yielding a first correction to it. In order to arrive at the full correction, we taken also into account modifications in the solid angle subtended by the atom due to the curvature of the spacetime. When the results found here are applied to a compact object with radius measuring three times the correspondent Schwarzschild radius ({\it i.e.}, like a neutron star), the black-body potential gravitationally corrected is almost two times lower than the flat case at $r=1.5R$ (or $r=4.5r_S$), corresponding to a force $70\%$ stronger. For the Sun, we verified that the Einstein's gravity reduces the black-body potential about $0.0004$ per cent in relation to this quantity when calculated in the flat spacetime at the same distance, which is therefore a very tiny difference when compared to the previous case.

We also have studied the black-body potential generated by a global monopole, which presents a deficit in the central solid angle associated to its spacetime. In opposition to the previous situation, we have found that the difference from the flat black-body potential grows with the radial distance, reaching a non-zero value in the infinity.

The analysis continued by taking into account the radiation emitted from an infinitely long cylinder, and we have calculated the attractive potential on the hydrogen atom related to the mentioned effect, taking into account just the source's shape. In this case, the BBF is attractive, falling with a square-inverse law at large distances, differently from that happens with spherical black bodies, in which the force obeys the inverse cube law at those distances as shown in \cite{Marte}. We must point out, as awaited, that the proper geometry of the body plays a central role in the BBF.

Next, we have appropriately introduced the angular parameter $\nu$ in the expression correspondent to the attractive potential due to the cylindric black-body radiation. This parameter multiplies the azimuthal (or central) angle which enters in the calculation of the solid angle subtended by the atom, characterizing the conical geometry of the cosmic string (or of its realization as a disclination in a crystal) spacetime, and thus the topological properties of this latter. The BBF vanishes when we consider an ideal static cosmic string without thickness. For the cosmic string spacetime the BBF correction is negligible since $\nu\approx 1$, but for topological defects in condensed matter this parameter can have a value such that the BBF correction becomes quite important \cite{Modes}.

\section*{Acknowledgments}

The authors acknowledge the financial support provided by Funda\c c\~ao Cearense de Apoio ao Desenvolvimento Cient\'\i fico e Tecnol\'ogico (FUNCAP), by  Conselho Nacional de Desenvolvimento Cient\'\i fico e Tecnol\'ogico (CNPq), and  FUNCAP/CNPq/PRONEX.

\end{document}